# Sensitive optical coating defect detection via second harmonic generation

Authors: J. Lukeš[1,2], V. Hájková[3], M. Hlubučková[1], V. Kanclíř[1,2] and K. Žídek[1]

[1] Institute of Plasma Physics of the Czech Academy of Sciences, Za Slovankou 1782/3, 182 00 Prague 8 - Liben, Czech Republic

[2] Faculty of Mechatronics, Informatics and Interdisciplinary Studies, Technical University of Liberec, Studentská 1402/2, 461 17, Liberec, Czech Republic

[3] Department of Radiation and Chemical Physics, Institute of Physics of the Czech Academy of Sciences, Na Slovance 1999/2, 182 20 Prague 8, Czech Republic

**Optical coating, an integral part of many optical systems, is prone to damage from environmental exposure and laser irradiation. This underscores the need for reliable and sensitive coating diagnostics. We introduce second harmonic generation (SHG) as a method for the sensitive detection of defects and inhomogeneities within optical thin films. We demonstrate the use of SHG on $Si_3N_4$ layers tested for their laser-induced damage threshold. The SHG mapping was compared with commonly used diagnostic techniques, including Nomarski microscopy, white light interferometry, and electron microscopy. Owing to its sensitivity to variations in local symmetry, SHG is able to discern minute alterations in material composition, mechanical stress, and interface structures. Therefore, SHG identified modifications in the tested layers, highly extending the damage recognized by standard methods. Furthermore, we demonstrate SHG's ability to enhance specific features by modulation of incidence angles and polarization modes. Based on these findings, we propose SHG as an ideal diagnostic tool for the early identification of laser-induced modifications in centrosymmetric thin films.**

Optical coating – a stack of interference thin films providing a targeted optical response of an optical element - is typically the most vulnerable part of optical systems. It is prone to damage by ambient environment or laser radiation. Since coating deposition is a complex process governed by many factors, even slight variation in the process or subtle contamination leads to notable differences in the coating properties and durability. [1] Therefore, ongoing research focuses on the efficient evaluation of coating properties with respect to the presence of defects or local inhomogeneities. [2–6] The defects may be triggered by the process of manufacturing optics (grinding, polishing, etc.)[6–8], by the deposition process[6,9,10], or even by the surrounding environment and optics handling. [6,11,12]

The defects have a major impact on the applicability of the coating in demanding applications, such as high-power lasers. One of the decisive aspects of laser coatings is their laser-induced damage threshold (LIDT). The threshold is determined by exposing the coating to high fluences of laser irradiation on predefined spots, and potential damage is evaluated afterward. Testing is typically done with a pulsed laser, and each spot on the sample is irradiated either with one pulse (1-on-1 test) or multiple pulses (s-on-1). Despite using low-repetition lasers, s-on-1 tests typically give significantly lower LIDT compared to the case when the layer is exposed to only one pulse.[13] Coatings undergo subtle, cumulative changes in the layer during the irradiation, which starts to be apparent only after repeated exposure to the laser beam. The ability to detect these subtle layer changes even before they become a discernible defect would be highly beneficial in studying the origin of the changes and their elimination.

In this letter, we report on the possibility of using second-harmonic generation (SHG) as a highly sensitive approach to non-destructively detect defects and inhomogeneities in thin films. The SHG phenomenon is very sensitive to variations in the local symmetry of a material. This provided us with the possibility to investigate very subtle changes in the layer that have a minimal effect on the linear optical response, such as the refractive index of the layers and the connected coating reflectance.[14] We demonstrate that the sensitivity of SHG outcompetes these commonly available methods, including optical microscopy, white light interferometry, differential interference contrast (DIC) microscopy, or electron microscopy.

In particular, we used SHG for detecting defects in Si3N4 layers induced by a focused laser beam. We show that the polarization and incident-angle-dependent SHG is an option to distinguish different regions of the damaged layers and open the way to potentially identify the source of changes for simple single-layer systems.

Measurements were carried out on several **samples** of single-layer $Si_3N_4$ coatings deposited with the RF magnetron sputtering from the Si target on the N-BK7 substrate – magnetron power was kept at 250 W for all depositions, and deposition pressure varied from 0.2 to 0.4 Pa in a nitrogen atmosphere. The samples differed only in their thickness, which ranged from 250 to 1000 nm. In this letter, we focus on two samples with thicknesses of 400 nm (Sample A) and 300 nm (Sample B). The samples were thoroughly characterized via transmission and reflectance spectroscopy, as well as ellipsometry.[15]

The samples were tested for their **laser-induced damage threshold (LIDT)**, which induced a number of defects over the sample, which could be reproducibly located with each characterization method. We used nanosecond Nd:YAG laser at 1064 nm with 10 ns pulses in the so-called 1-on-1 configuration with pulse energies ranging from 5-15 mJ.[13] The pulses were focused into a spot with a diameter of approx. 200 μm. The focal spot shape was irregular, as is apparent from the figures below. Nevertheless, we used the laser only to induce the studied defects.

The **SHG measurements** were performed with a setup described elsewhere[16]. The laser used in the setup was a pulsed femtosecond laser (215 fs pulses) with a wavelength of 1028 nm, 100 kHz rep. rate and irradiation peak power 400 GW/cm$^2$ focused into a Gaussian spot with the FWHM of 20 μm. The setup enabled the SHG mapping of samples in the reflection geometry. The mapping was carried out for varying incident angles and polarization of fundamental and SHG beams. During the SHG mapping, we focused only on the probed LIDT sites. This was convenient to reduce the acquisition time, as the acquisition of a 50x50 pixel (1x1 mm) map typically took 4 hours. It is, however, worth noting that this acquisition speed results from point-by-point scanning, and it can be decreased by several orders of magnitude by using sensitive array detectors.

Our layers of Si3N4 were centrosymmetric and, therefore, did not inherently generate the strong dipolar bulk SH induced by an electric field. Nevertheless, weak SHG is observed due to quadrupole electric field contributions, defects, and local symmetry breaking due to residual mechanical strain, inhomogeneities, and the interfaces themselves[16–18].

In addition, other **commonly used tested methods** were used to compare the sensitivity of each method. We tested the use of: (i) electron microscopy (TESCAN VEGA3) in backscattered electrons (BSE) regime with an accelerating voltage of 20 kV (for sample A) and 3 kV (for sample B); (ii) a microscope with differential interference contrast (DIC) capability Olympus BX51M with halogen illumination, and Keyence VHX-7000N with LED illumination; (iii) white light interferometer ZYGO NV 7200.

This work mainly focused on studying the damaged or modified spots of optical thin films via SHG. The aim was not only to determine whether the spot was damaged but also to examine the extent of the damage and the adjacent area of the "burnt" spot. Therefore, the studied areas were selected to include visually damaged spots, as well as the spots corresponding to pulse energies close to the LIDT, where a standard microscope did not observe any damage. We will use the term "burnt" spot for the area where the local transmittance and reflectance were clearly changed.

In Figure 1, upper line (sample A), we provide an example of a damaged spot apparent to the naked eye. The ellipse-shaped burnt area in the middle corresponds to a delaminated and melted layer. The spot was investigated using various methods, and the corresponding maps were cropped and scaled to show the same area. For scanning electron microscopy (SEM) and white light interferometry (WLI) the central spot is clearly visible, while no changes are visible in the surrounding area around the damaged spot. This arises due to the strong sensitivity of the methods on the surface chemical properties for SEM and morphology for WLI. Both these properties were modified in the burnt central part, while stayed the same outside of this area.

Differential interference contrast microscopy (DIC) shows changes in the immediate vicinity of the spot, above and below, where there are "brighter" areas. The color variation represents a change in the optical path and, therefore, a change in the thickness or refractive index of the layer. Such changes may be caused, for example, by changes in mechanical stress in the layer connected with subtle variations in refractive index and thickness[19,20] or surface topography[21,22]. Images from DIC for both samples in Figure 1 were heavily adjusted in terms of contrast and color balance to highlight changes around burnt spots.

Finally, we investigated the spot using SHG at a high incident angle of 70 deg. and pP polarization configuration, which denotes the p-polarized incident beam and P-polarized measured SHG – see Figure 1A, SHG panel. The burnt area itself is prominent with a very low SHG efficiency (dark blue part). SHG intensity for this feature was decreased by 60% compared to the pristine layer. Nevertheless, it is surrounded by an extensive damaged area with a decreased SHG by 30% compared to the pristine layer (lighter blue and green area). This area extended hundreds of microns beyond the apparent damage. According to the SHG intensity, the area of laser-induced changes in the layer is much larger than previous methods suggest, and the changes gradually transition to a pristine layer SHG.

The damage on the second example spot (Figure 1B) is subtle but again shows a trend similar to the previous case. The most notable features are burnt spots, which correspond to local imperfections – absorbing spots – which were placed in the focal laser spot center. SEM and the WLI are again sensitive to the burnt spots themselves. A severe problem for SEM for both samples was the local charging, which obscured the measured maps. For instance, the brighter "shadow" below the spot for Sample A is an artifact induced by charging. Similarly, a brighter area around two spots on sample B might be caused by charging, or it can arise from real changes in the sample since the shape of the area resembles the area measured with SHG. In this case, the charging artefacts disqualified SEM from providing reliable interpretation.

The DIC microscopy is able to detect small changes around the two burnt spots extending tens of microns. Note that the gradual shift in color observable across the whole image can be ascribed to the image vignetting, i.e. uneven light collection efficiency. Nevertheless, also here the SHG image (pP pol., 70 deg.) shows superior sensitivity, as the variation in SHG can be observed in the burnt spots (dark areas) but also in the area extending almost 200 μm far from the burnt spots.

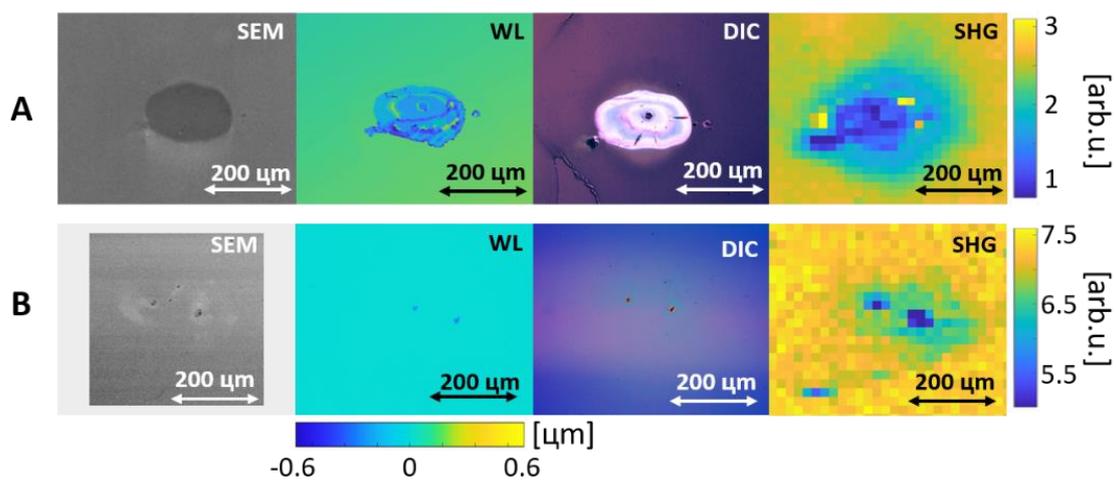

*Figure 1 – Comparison of different methods for imaging of LIDT damage – **SEM** (Scanning electron microscopy) in BSE mode; **WL** (White light interferometry); **DIC** (Differential interference contrast microscopy); **SHG** (Second-harmonic generation method) 70 deg. pP pol. Laser: λ = 1028 nm. All maps use the same scale. Panels A: Sample A: $Si_3N_4$ layer (400 nm); Panels B: Sample B: $Si_3N_4$ layer (300 nm). Both at N-BK7 substrate*

Figure 1 demonstrates the sensitivity of SHG for a certain experimental configuration. Nevertheless, SHG in thin films originates from several sources, such as interfaces, internal mechanical stress, or bulk quadrupolar interaction. Their proportion can be tuned by changing both the polarization and the incident angle. Therefore, we studied the polarization dependence of SHG from the damaged sites – see Figure 2. The measurements were performed with an incidence angle of 30 deg. and in three polarization configurations – pP, sS, and mixS, where "mix" indicates the equal contribution from p and s polarization. This measurement was carried out for a highly damaged spot, where the SHG in the pP configuration (see Figure 2A) was able to track extensive changes in the layer far beyond the visibly damaged area.

The materials with ∞m symmetry, which is the case for both the pristine Si3N4 layers, the N-BK7 substrate, and all involved interfaces, are expected to emit the strongest SHG in the pP configuration and none in the sS configuration[23]. In accordance with the expectations, the pristine layer featured low SHG in the "sS" configuration (mean int. 0.7 arb.u.), and significant SHG in the "mixS" (2.5 a.u.) and "pP" (7 a.u.) configurations. Small SHG in the "sS" configuration can be explained by residual mechanical stress and layer microstructure, which both elevate the local symmetry.

The "pP" configuration has proved to be the most sensitive case, where the SH intensity was altered both in the damaged spot itself and in its vicinity. The s-polarized incident beam promotes SHG from the visibly damaged spot with a very high contrast. This can be ascribed to the changed local symmetry in the burnt spot. Note that the bright SHG spot in the upper part of the images in Fig. 2 is an example of an SHG signal of a particle attached to the layer surface. We verified this by remeasuring the same area after the cleaning procedure.

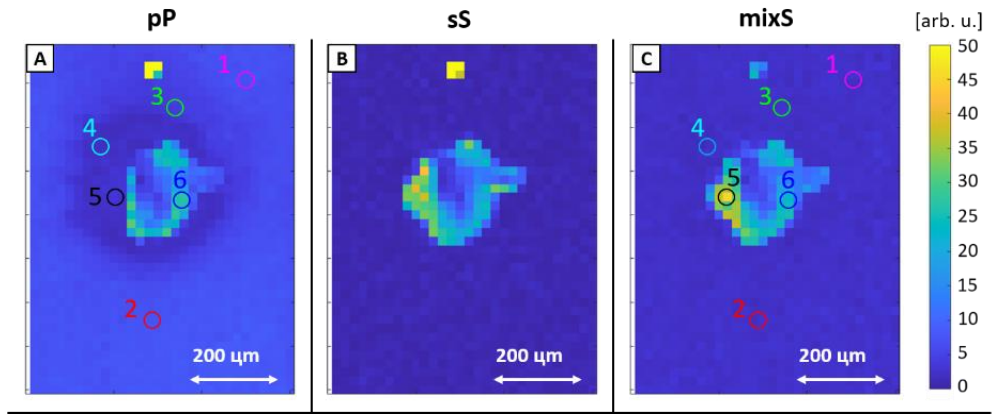

*Figure 2 – Polarization dependence of SHG – an area of the damaged spot was observed under different polarization configurations (pP – A, sS – B, and mixS – C), incident angle 30 deg. Six selected sites (marked by circles in panels) were measured under different incident angles – see Figure 3. Sample A: $Si_3N_4$ layer (400 nm) @ N-BK7 substrate. Laser: $\lambda$ = 1028 nm.*

More information about the damaged spot can be attained from the dependency of SHG on the incident angle. We provide the information for six selected spots from Fig. 2, which correspond to the pristine layer, burnt spot, and modified layer. We measured the incident angle dependence for "pP" and "mixS" polarizations for these spots. The resulting angular dependences in Fig. 3 were normalized to allow for a better comparison between the shapes. For two spots in the pristine layer surrounding the damage (**Points 1 and 2**), SHG angular dependences overlap perfectly. This indicates that the layer in these areas was not changed and illustrates the noise level in this measurement. **Points 3 and 4**, which were in the near vicinity of the burnt spot, have in the "mixS" configuration (Figure 3 B) the same angle dependency as the pristine layer, although in the "pP" configuration, changes are more noticeable. This behavior corresponds to the case where the overall symmetry is conserved, while the contribution of different SHG sources varies and alters the angular SHG dependence. **Points 5 and 6**, which were in the burnt spot itself, have largely different angle dependencies compared to the pristine layer - both in amplitude and shape. That means that the local morphology of the layer is entirely different. We can speculate that this results from local inhomogeneities, which also led to the light scattering.

Finally, we used the SHG dependence on incident angle Θ to quantify the observed changes in the layer. We selected a spot on the layer edge as a reference $I_{11}(\theta)$ and calculated the total SHG angular dependence variation σ for each pixel {i,j}: $\sigma_{ij} = \sum_\theta \bigl(I_{ij}(\theta)/\langle I_{ij}(\theta)\rangle - I_{11}(\theta)/\langle I_{11}(\theta)\rangle\bigr)^2$. To focus on the shape variations, we normalize each pixel dependence by its mean value. As a result, the low variation values corresponded to a similar shape of incident angle dependence – see Fig. 3 C for the "pP" configuration. We observed that concentric areas around the epicenter of the burnt spot feature similar angle dependence of SHG, which corresponds well with the expected distribution of the laser energy.

While this map was very useful to identify areas with an analogous laser-induced modification, it should be evaluated together with the corresponding SHG intensity map (see Fig. 3A). For instance, the dark "ring" around the burnt spot in Fig. 3C does not imply a lack of laser-induced damage. Instead, it corresponded to changes, which led in this particular configuration to a similar angular dependence, while showing decreased SHG intensity.

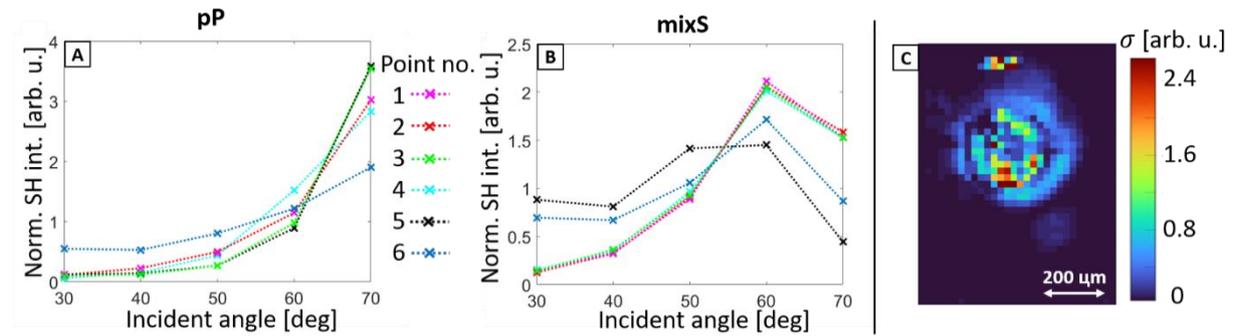

*Figure 3 – A,B: Incident angle dependence for marked spots in Fig.2 (A: pP polarization, B: mixS polarization), intensities were separately normalized by their mean values; C: map of SHG angular dependence variation for "pP" pol. scheme – see text for σ calculation. Sample A: $Si_3N_4$ layer (300 nm) @ N-BK7 substrate. Laser: $\lambda$ = 1028 nm.*

Overall, by combining the polarization and angle-dependent measurements, we are able to discern areas with distinct behavior – see Figures 2 and 3. The internal damaged part features entirely different SHG properties – likely due to the remaining melted layer material with an entirely different morphology from the original layer. The area surrounding the central damaged spot can be ascribed to the layer, which went through a heating cycle but was only partly modified. Analogously to annealing, laser-induced heating can lead to changes in internal stress[24] and layer microstructure[25]. As such, the SHG will feature the same polarization properties and similar angle dependence, but the total intensity will vary. Potentially, differences in the angle dependence can be interpreted by a theoretical model in terms of surface and bulk SHG contribution and test various models of changes in the layer.

We propose SHG to be an ideal, highly sensitive tool to study laser-induced layer modification. We demonstrated the high sensitivity of the method on selected damaged spots of Si3N4, which are typical examples of many studied sites. SHG can reveal subtle variations in the layer well before its apparent damage is visible in other standard methods, such as DIC, and before the layer variations lead to the loss of the layer optical properties.

**Acknowledgment**

This work was co-funded by the European Union and the state budget of the Czech Republic under the project LasApp CZ.02.01.01/00/22_008/0004573. We also gratefully acknowledge financial support of Student Grant Competition at the Technical University of Liberec (under the project No SGS-2024-3419).

1.	Stenzel, O., Harhausen, J., Gäbler, D., Wilbrandt, S., Franke, C., Foest, R. & Kaiser, N. Investigation on the reproducibility of optical constants of $TiO_2$, $SiO_2$, and $Al_2O_3$ films, prepared by plasma ion assisted deposition. *Opt. Mater. Express, OME* **5**, 2006–2023 (2015).

2.	Bu, C., Li, R., Liu, T., Shen, R., Wang, J. & Tang, Q. Micro-crack defects detection of semiconductor Si-wafers based on Barker code laser infrared thermography. *Infrared Physics & Technology* **123**, 104160 (2022).

3.	Shang, X., Shi, W., Su, J. & Dong, C. Study on the laser-induced damage of thin films by terahertz time-domain spectroscopy. *Front. Phys.* **10**, (2022).

4.	Taherimakhsousi, N., MacLeod, B. P., Parlane, F. G. L., Morrissey, T. D., Booker, E. P., Dettelbach, K. E. & Berlinguette, C. P. Quantifying defects in thin films using machine vision. *npj Comput Mater* **6**, 1–6 (2020).

5.	Zhu, J., Liu, J., Xu, T., Yuan, S., Zhang, Z., Jiang, H., Gu, H., Zhou, R. & Liu, S. Optical wafer defect inspection at the 10 nm technology node and beyond. *Int. J. Extrem. Manuf.* **4**, 032001 (2022).

6.	Dong, S., Jiao, H., Wang, Z., Zhang, J. & Cheng, X. Interface and defects engineering for multilayer laser coatings. *Progress in Surface Science* **97**, 100663 (2022).

7.	Shaw-Klein, L. J., Burns, S. J. & Jacobs, S. D. Model for laser damage dependence on thin-film morphology. *Appl. Opt., AO* **32**, 3925–3929 (1993).

8.	Bloembergen, N. Role of Cracks, Pores, and Absorbing Inclusions on Laser Induced Damage Threshold at Surfaces of Transparent Dielectrics. *Appl. Opt., AO* **12**, 661–664 (1973).

9.	Kaiser, N. Review of the fundamentals of thin-film growth. *Appl. Opt., AO* **41**, 3053–3060 (2002).

10.	Mustafa, M. K., Majeed, U. & Iqbal, Y. Effect on Silicon Nitride thin Films Properties at Various Powers of RF Magnetron Sputtering. *International Journal of Engineering & Technology* **7**, 39–41 (2018).

11.	Ames, D. P. & Chelli, S. J. Surface contamination effects on film adhesion on metals and organic polymers. *Surface and Coatings Technology* **187**, 199–207 (2004).

12.	Blech, I. & Cohen, U. Effects of humidity on stress in thin silicon dioxide films. *Journal of Applied Physics* **53**, 4202–4207 (1982).

13.	Pakalnytė, R., Pupka, E. & Melninkaitis, A. Direct comparison of laser-induced damage threshold testing protocols on dielectric mirrors: effect of nanosecond laser pulse shape at NIR and UV wavelengths. in *Laser-induced Damage in Optical Materials 2019* vol. 11173 92–101 (SPIE, 2019).

14.	Lukeš, J., Kanclíř, V., Václavík, J., Melich, R., Fuchs, U. & Žídek, K. Optically modified second harmonic generation in silicon oxynitride thin films via local layer heating. *Sci Rep* **13**, 8658 (2023).

15.	Kanclíř, V., Václavík, J. & Žídek, K. Precision of Silicon Oxynitride Refractive-Index Profile Retrieval Using Optical Characterization. *Acta Phys. Pol. A* **140**, 215–221 (2021).

16.	Das, N. K., Kanclíř, V., Mokrý, P. & Žídek, K. Bulk and interface second harmonic generation in the $Si_3N_4$ thin films deposited via ion beam sputtering. *J. Opt.* **23**, 024003 (2021).

17.	Gielis, J. J. H., Gevers, P. M., Aarts, I. M. P., van de Sanden, M. C. M. & Kessels, W. M. M. Optical second-harmonic generation in thin film systems. *Journal of Vacuum Science & Technology A* **26**, 1519–1537 (2008).

18.	Koskinen, K., Czaplicki, R., Kaplas, T. & Kauranen, M. Recognition of multipolar second-order nonlinearities in thin-film samples. *Opt. Express, OE* **24**, 4972–4978 (2016).

19.	Chen, Z., Tian, Y., Zhu, J., Sun, L., Wang, F., Ai, Y., Liu, H., Deng, X., Chen, M., Cheng, J. & Zhao, L. Laser Damage Performance Study of Fundamental Frequency Dielectric Film Optical Elements. *Crystals* **13**, 571 (2023).

20. Liu, H., Wang, L., Jiang, Y., Li, S., Liu, D., Ji, Y., Zhang, F. & Chen, D. Study on SiO2 thin film modified by post hot isostatic pressing. *Vacuum* **148**, 258–264 (2018).

21. Sozet, M., Néauport, J., Lavastre, E., Roquin, N., Gallais, L. & Lamaignère, L. Laser damage density measurement of optical components in the sub-picosecond regime. *Opt. Lett., OL* **40**, 2091–2094 (2015).

22. Kang, M. J., Park, T. S., Kim, M., Hwang, E. S., Kim, S. H., Shin, S. T. & Cheong, B.-H. Periodic surface texturing of amorphous-Si thin film irradiated by UV nanosecond laser. *Opt. Mater. Express, OME* **9**, 4247–4255 (2019).

23. Heinz, T. F. Chapter 5 - Second-Order Nonlinear Optical Effects at Surfaces and Interfaces. in *Modern Problems in Condensed Matter Sciences* (eds. Ponath, H.-E. & Stegeman, G. I.) vol. 29 353–416 (Elsevier, 1991).

24. Fourrier, A., Bosseboeuf, A., Bouchier, D. B. D. & Gautherin, G. G. G. Annealing Effect on Mechanical Stress in Reactive Ion-Beam Sputter-Deposited Silicon Nitride Films. *Jpn. J. Appl. Phys.* **30**, 1469 (1991).

25. Huang, L., Ding, Z., Yuan, J., Zhou, D. & Yin, Z. Effect of the post-heating temperatures on the microstructure, mechanical and electrical properties of silicon nitride thin films. *Ceramics International* **48**, 9188–9196 (2022).

**Conflict of interest**

The authors have no conflicts to disclose.

**Author Contributions**

**Lukeš**: Data Curation (lead), Formal Analysis (lead), Investigation (lead), Methodology (equal), Writing (equal)**, Hájková:** Investigation (supporting)**, Hlubučková:** Investigation (supporting)**, Kanclíř:** Resources (lead)**, Žídek**: Conceptualization (lead), Formal Analysis (supporting), Funding Acquisition (lead), Methodology (equal), Supervision (lead), Writing (equal)

**Data Availability**

The data that support the findings of this study are available from the corresponding author upon reasonable request.